\documentclass{ws-ijmpd}
\usepackage[super]{cite}
\usepackage{graphicx}
\begin{document}

\markboth{H. S. VIEIRA et al.}
{Quantum relativistic cosmology}

\catchline{}{}{}{}{}

\title{Quantum relativistic cosmology: dynamical interpretation and tunneling universe}

\author{H. S. VIEIRA}
\address{Departamento de F\'{i}sica, Universidade Federal da Para\'{i}ba, Caixa Postal 5008, CEP 58051-970, Jo\~{a}o Pessoa, PB, Brazil\\
Instituto de F\'{i}sica de S\~{a}o Carlos, Universidade de S\~{a}o Paulo, Caixa Postal 369, CEP 13560-970, S\~{a}o Carlos, S\~{a}o Paulo, Brazil\\
horacio.santana.vieira@hotmail.com}

\author{V. B. BEZERRA}
\address{Departamento de F\'{i}sica, Universidade Federal da Para\'{i}ba, Caixa Postal 5008, CEP 58051-970, Jo\~{a}o Pessoa, PB, Brazil\\
valdir@fisica.ufpb.br}

\author{C. R. MUNIZ}
\address{Grupo de F\'isica Te\'orica (GFT), Universidade Estadual do Cear\'a, Faculdade de Educa\c c\~ao, Ci\^encias e Letras de Iguatu, Iguatu, Cear\'a, Brazil\\
celio.muniz@uece.br}

\author{M. S. CUNHA}
\address{Grupo de F\'isica Te\'orica (GFT), Centro de Ci\^encias e Tecnologia, Universidade Estadual do Cear\'a, CEP 60714-903, Fortaleza, Cear\'a, Brazil\\
marcony.cunha@uece.br}

\maketitle

\begin{history}
\received{Day Month Year}
\revised{\today}
\end{history}

\begin{abstract}
In this work the wave functions associated to the quantum relativistic universe, which is described by the Wheeler-DeWitt equation, are obtained. Taking into account different kinds of energy density, namely, matter, radiation, vacuum, dark energy, and quintessence, we discuss some aspects of the quantum dynamics. In all these cases, the wave functions of the quantum relativistic universe are given in terms of the triconfluent Heun functions. We investigate the expansion of the universe using these solutions and found that the asymptotic behavior for the scale factor is $a(t) \sim \mbox{e}^{t}$ for whatever the form of energy density is. On the other hand, we analyze the behavior at early stages of the universe and found that $a(t) \sim t^{1/2}$. We also calculate and analyze the transmission coefficient through the effective potential barrier.

\keywords{Wheeler-DeWitt equation; triconfluent Heun equation; energy density; boundary condition}
\end{abstract}

\ccode{PACS Nos.: 02.30.Gp, 03.65.Ge, 04.20.Jb, 04.60.Ds, 04.60.Pp, 04.62.+v, 95.30.Sf, 98.80.Qc, 98.80.Jk}

%
%
\section{Introduction}
In the latter 1950's Peres and Rosen \cite{NuovoCimento.13.430} studied the problem concerning the Cauchy's conditions which should be satisfied by Einstein's equations. A few years later, Peres \cite{NuovoCimento.26.53} continued to analyze the question related to the construction of initial data for General Relativity and in this context developed the Hamilton-Jacobi formalism for an arbitrary gravitational field.

The Wheeler-DeWitt equation \cite{Wheeler:1968,PhysRev.160.1113} was considered at that time as the equation which correctly describes quantum gravity phenomena (For a discussion about different aspects of the Wheeler-DeWitt equation, see Rovelli paper \cite{ClassQuantumGrav.32.124005}).

In fact, the Wheeler-DeWitt equation is not the equation that describes the behavior of the spacetime from the quantum point of view, but it has offered the first conceptual approach which combined quantum mechanics and general relativity. Additionally, it has played fundamental role in the development of the ideas concerning the formalism of a possible quantum theory of the gravitational field \cite{arXiv:0909.2566,arXiv:gr-qc/0101003,arXiv:0804.0672}.

Since the early 1960's up to now, different investigations have been done, as for example in quantum cosmology, in which context the wave function that describes the quantum evolution of the universe is obtained by using the Wheeler-DeWitt Equation \cite{PhysRevD.86.063504,ClassQuantumGrav.30.143001,PhysRevD.89.083510,PhysRevD.89.043526,PhysRevD.99.066010}.

The Wheeler-DeWitt equation was also discussed and generalized in different contexts as for example in Eddington inspired Born-Infeld Gravity \cite{JCAP.1812.032}, in Bianchi scalar-field cosmology \cite{EurPhysJC.76.225}, in the brane approach \cite{PhysRevD.77.066017} as a possible mechanism to mimic the effects of dark energy \cite{PhysLettB.661.37}, among others. Exact solutions of the Wheeler-DeWitt equation were also obtained and their possible consequences studied. Along this line of research there is an extensive literature, from which we mention \cite{ClassQuantumGrav.11.1211,PhysLettB.725.463,AnnPhys.359.80,GenRelGrav.48.13}.

Therefore, taking into account that the Wheeler-DeWitt equation gives us a possible path to understand aspects of the construction of an eventual quantum theory of gravity, and that along the last almost six decades it has inspired a lot of researches, it seems to be important to find solutions of this equation and use them to study some aspects of the quantum dynamics, as well as to study many others phenomena.

Recently, we used the description of the universe constructed in the framework of classical mechanics, assuming the cosmological principle and the fact that the universe experiences an expansion. Then, we analyzed, in this Newtonian description of the universe, some aspects of quantum cosmology by considering that the wave functions which describe the evolution of the universe are given by the solutions of the Schr\"{o}dinger equation \cite{JMathPhys.60.102301}.

In this work, we extended this analysis in the sense that now we consider the relativistic context. Thus, we determine the solutions of the Wheeler-DeWitt equation and study some of their consequences instead of using the Schr\"{o}dinger equation as was done recently in the context of Newtonian cosmology. These solutions are given in terms of the triconfluent Heun functions. The Heun's functions have been recently used in the context of quantum relativistic cosmology, as for example, in the resolution of type IV singularities in quantum cosmology \cite{PhysRevD.89.064016} and to investigate the Wheeler-DeWitt equation for hyperbolic universes \cite{IntJTheorPhys.57.652}, among others.

This paper is organized as follows. In Sec. \ref{Sec.II}, we write the Wheeler-DeWitt equation in the Friedmann-Robertson-Walker (FRW) universe. In Sec. \ref{Sec.III}, we analyze the dy\-nam\-i\-cal in\-ter\-pre\-ta\-tion. In Sec. \ref{Sec.IV}, we discuss the tunneling universe. Finally, in Sec. \ref{Sec.V}, we present our conclusions.
%
%
\section{Wheeler-DeWitt equation in the FRW universe}\label{Sec.II}
Let us start by assuming a recent result \cite{PhysRevD.94.023511} which shows that the classical evolution of the universe at different stages can be obtained from the following Hamiltonian
\begin{equation}
H(P_{a},a)=\frac{3 \pi c^{2}}{4G}a^{3}\left(\frac{4G^{2}P_{a}^{2}}{9 \pi^{2} c^{4} a^{4}}+\frac{k c^{2}}{a^{2}}-\frac{8 \pi G}{3 c^{2}}\rho\right),
\label{eq:Hamiltonian_WDW}
\end{equation}
where $P_{a}$ is the momentum, $a$ is the scale factor, such that $0 \leq a < \infty$, and $k=-1,0,+1$. This form of the Hamiltonian is equally valid for different kinds of energy density $\rho$. In order to write down the Wheeler-DeWitt equation in the minisuperspace approximation, we do the following replacement $P_{a}^{2} \rightarrow -\frac{\hbar^{2}}{a^{p}}\frac{\partial}{\partial a}\left(a^{p}\frac{\partial}{\partial a}\right)$, and then substitute the classical Hamiltonian constraint $H=0$ by $H\Psi(a)=0$, for the wave function associated to the gravitational field. The parameter $p$ represents the uncertainty in the ordering of factors $a$ and $\partial / \partial a$. For simplicity, we only consider the original undeformed momentum conjugate to the scale factor, that is, the case when $p=0$. Therefore, in the FRW universe, the Wheeler-DeWitt equation reads
\begin{equation}
\left\{\frac{d^{2}}{da^{2}}-\frac{9\pi c^{4}a^{2}}{4\hbar^{2}G^{2}}\left[kc^{2}-\frac{8\pi Ga^{2}}{3c^{2}}(\rho_{\omega}+\rho_{v})\right]\right\}\Psi(a)=0,
\label{eq:WDE}
\end{equation}
where we are assuming that the wave function depends only on the scale factor. The parameter $\rho_{v}$ is the energy density of the vacuum and $\rho_{\omega}$ represents the other kinds of energy which we will consider.

The energy density of the vacuum, $\rho_{v}$, can be expressed in terms of the cosmological constant, as follows:
\begin{equation}
\rho_{v}=\frac{\Lambda c^{4}}{8\pi G}.
\label{eq:WDE_energy_density_vacuum}
\end{equation}
The energy density parameter $\rho_{\omega}$, which represents different kinds of energy, namely, matter, radiation, dark energy, and quintessence, can be expressed as
\begin{equation}
\rho_{\omega}=\frac{A_{\omega}}{a^{3(\omega+1)}},
\label{eq:WDE_density}
\end{equation}
where
\begin{equation}
A_{\omega}=\rho_{\omega 0}a_{0}^{3(\omega+1)},
\label{eq:A_WDE_density}
\end{equation}
and $\rho_{\omega 0}$ stands for the value of $\rho_{\omega}$ at present time, with the state parameter, $\omega$, being given by
\begin{equation}
\omega=\left\{
\begin{array}{rl}
	0            & \mbox{for matter}\ (A_{m}=\rho_{m0}a_{0}^{3}),\\
	\frac{1}{3}  & \mbox{for radiation}\ (A_{r}=\rho_{r0}a_{0}^{4}),\\
	-\frac{1}{3} & \mbox{for dark energy}\ (A_{d}=\rho_{d0}a_{0}^{2}),\\
	-\frac{2}{3} & \mbox{for quintessence}\ (A_{q}=\rho_{q0}a_{0}).
\end{array}
\right.
\label{eq:omega_universe}
\end{equation}

Taking into account the contribution of matter, radiation, dark energy, and quintessence to the content of energy, then the total energy density $\rho$ can be written as the sum
\begin{equation}
\rho=\sum_{\omega}\rho_{\omega}=\rho_{m}+\rho_{r}+\rho_{d}+\rho_{q}.
\label{eq:density_sums}
\end{equation}
Substituting Eqs.~(\ref{eq:WDE_energy_density_vacuum})-(\ref{eq:density_sums}) into Eq.~(\ref{eq:WDE}), the following equation is obtained
\begin{equation}
-\hbar^{2}\frac{d^{2}\Psi(a)}{da^{2}}+V_{eff}(a)\Psi(a)=0,
\label{eq:WDE_mov_1}
\end{equation}
where $V_{eff}(a)$ is the effective potential, which is written as
\begin{equation}
V_{eff}(a)=\frac{9\pi c^{6}k}{4G^{2}}a^{2}-\frac{6\pi^{2}c^{2}}{G}\biggl(A_{r}+A_{m}a+A_{d}a^{2}+A_{q}a^{3}+\frac{\Lambda c^{4}}{8\pi G}a^{4}\biggr).
\label{eq:WDW_effective_potential_energy}
\end{equation}

Equation (\ref{eq:WDE_mov_1}) is similar to an one-dimensional time-independent Schr\"{o}dinger equation for a particle with $1/2$ of the unit mass and zero energy.

It is worth noticing that this expression given by Eq.~(\ref{eq:WDW_effective_potential_energy}) tell us that the effective potential can be positive or negative depending on the contributions of the different kinds of energy as well as on the value of the cosmological constant.

\begin{figure}[htbp]
	\centering
		\includegraphics[scale=0.50]{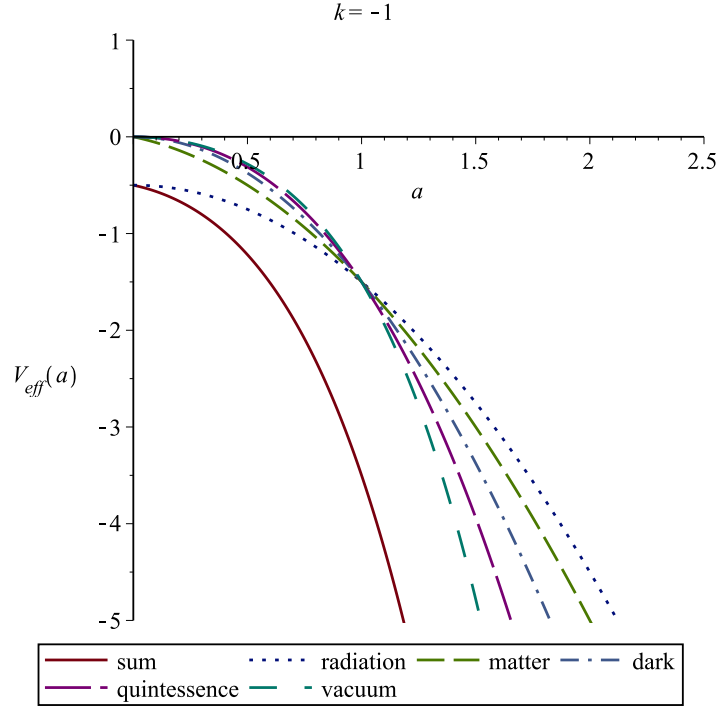}
	\caption{The effective potential energy, $V_{eff}(a)$, for $\Lambda > 0$. We focus on the $k=-1$ case and compare with each kind of energy density $\omega$.}
	\label{fig:WDW_Fig1}
\end{figure}

\begin{figure}[htbp]
	\centering
		\includegraphics[scale=0.50]{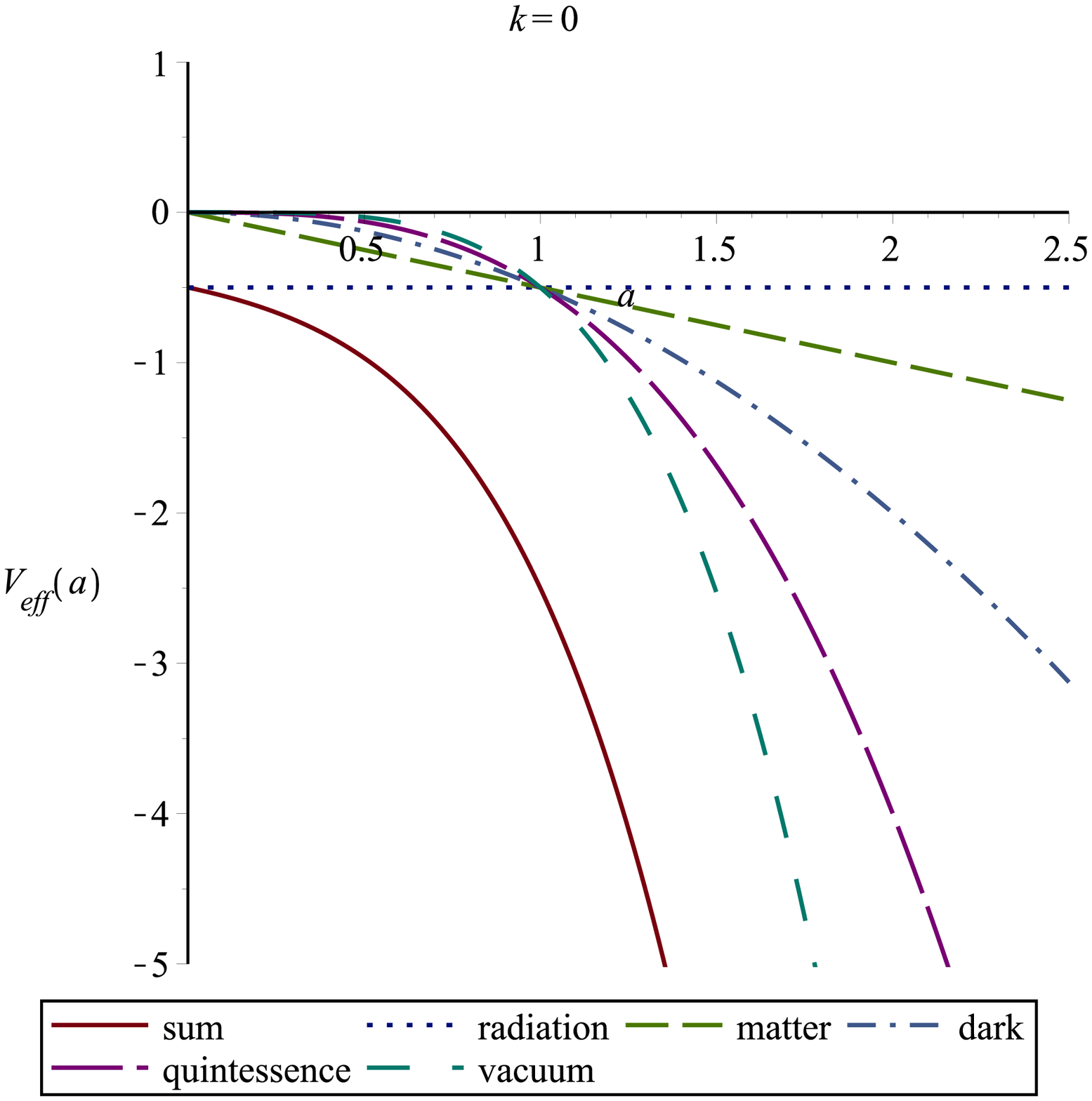}
	\caption{The effective potential energy, $V_{eff}(a)$, for $\Lambda > 0$. We focus on the $k=0$ case and compare with each kind of energy density $\omega$.}
	\label{fig:WDW_Fig2}
\end{figure}

\begin{figure}[htbp]
	\centering
		\includegraphics[scale=0.50]{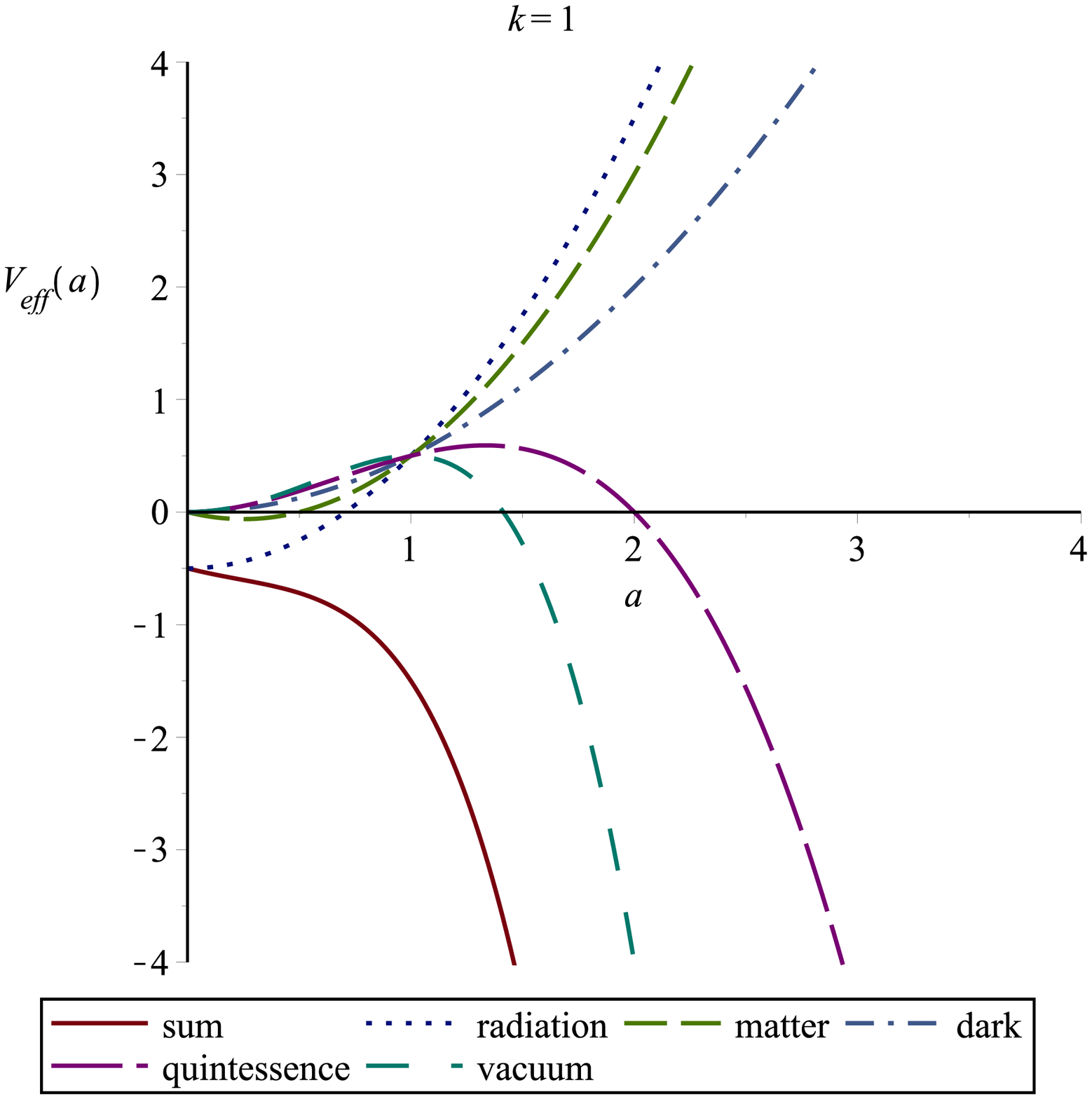}
	\caption{The effective potential energy, $V_{eff}(a)$, for $\Lambda > 0$. We focus on the $k=1$ case and compare with each kind of energy density $\omega$.}
	\label{fig:WDW_Fig3}
\end{figure}

\begin{figure}[htbp]
	\centering
		\includegraphics[scale=0.50]{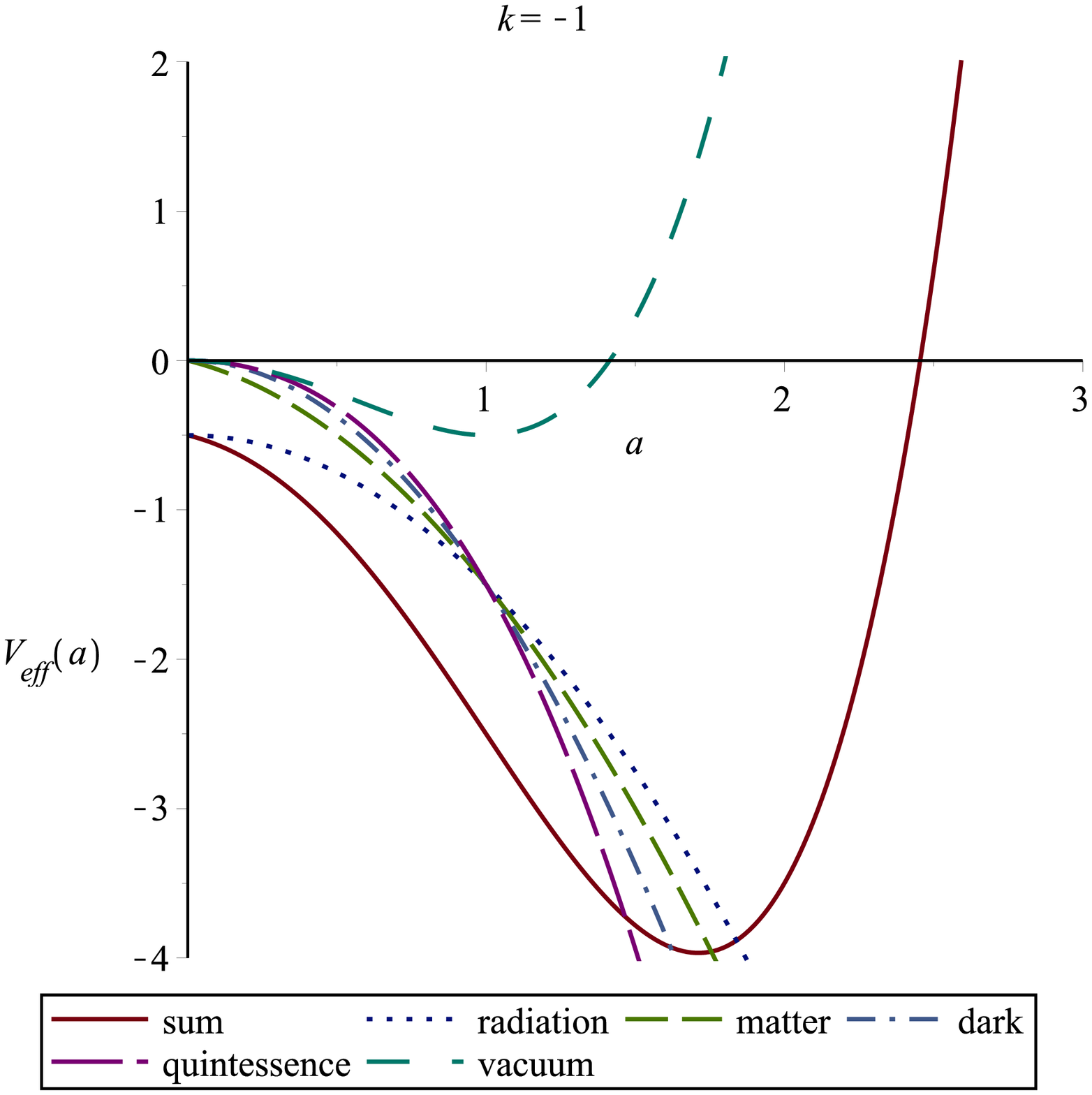}
	\caption{The effective potential energy, $V_{eff}(a)$, for $\Lambda = -|\Lambda|$. We focus on the $k=-1$ case and compare with each kind of energy density $\omega$.}
	\label{fig:WDW_Fig4}
\end{figure}

\begin{figure}[htbp]
	\centering
		\includegraphics[scale=0.50]{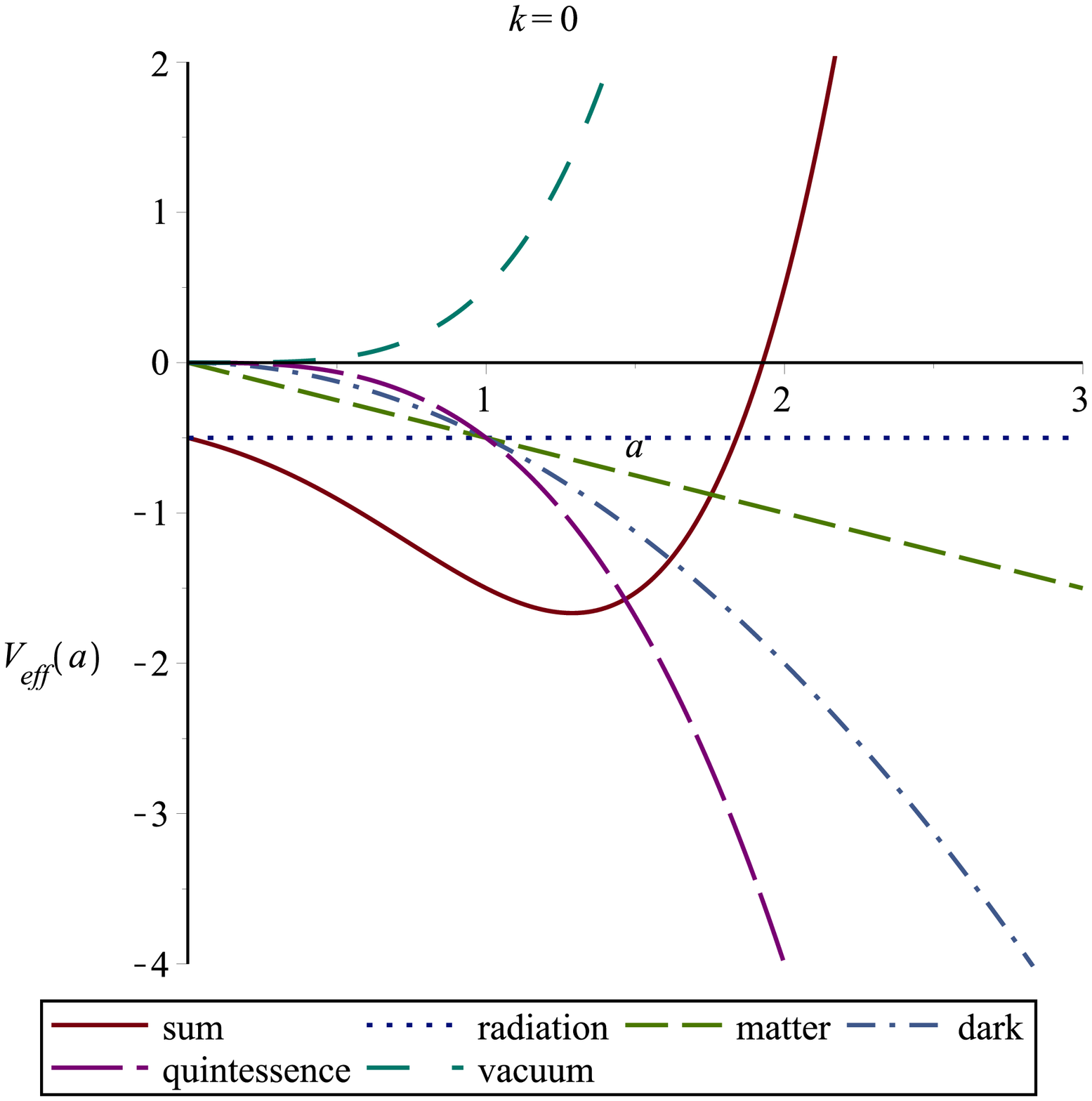}
	\caption{The effective potential energy, $V_{eff}(a)$, for $\Lambda = -|\Lambda|$. We focus on the $k=0$ case and compare with each kind of energy density $\omega$.}
	\label{fig:WDW_Fig5}
\end{figure}

\begin{figure}[htbp]
	\centering
		\includegraphics[scale=0.50]{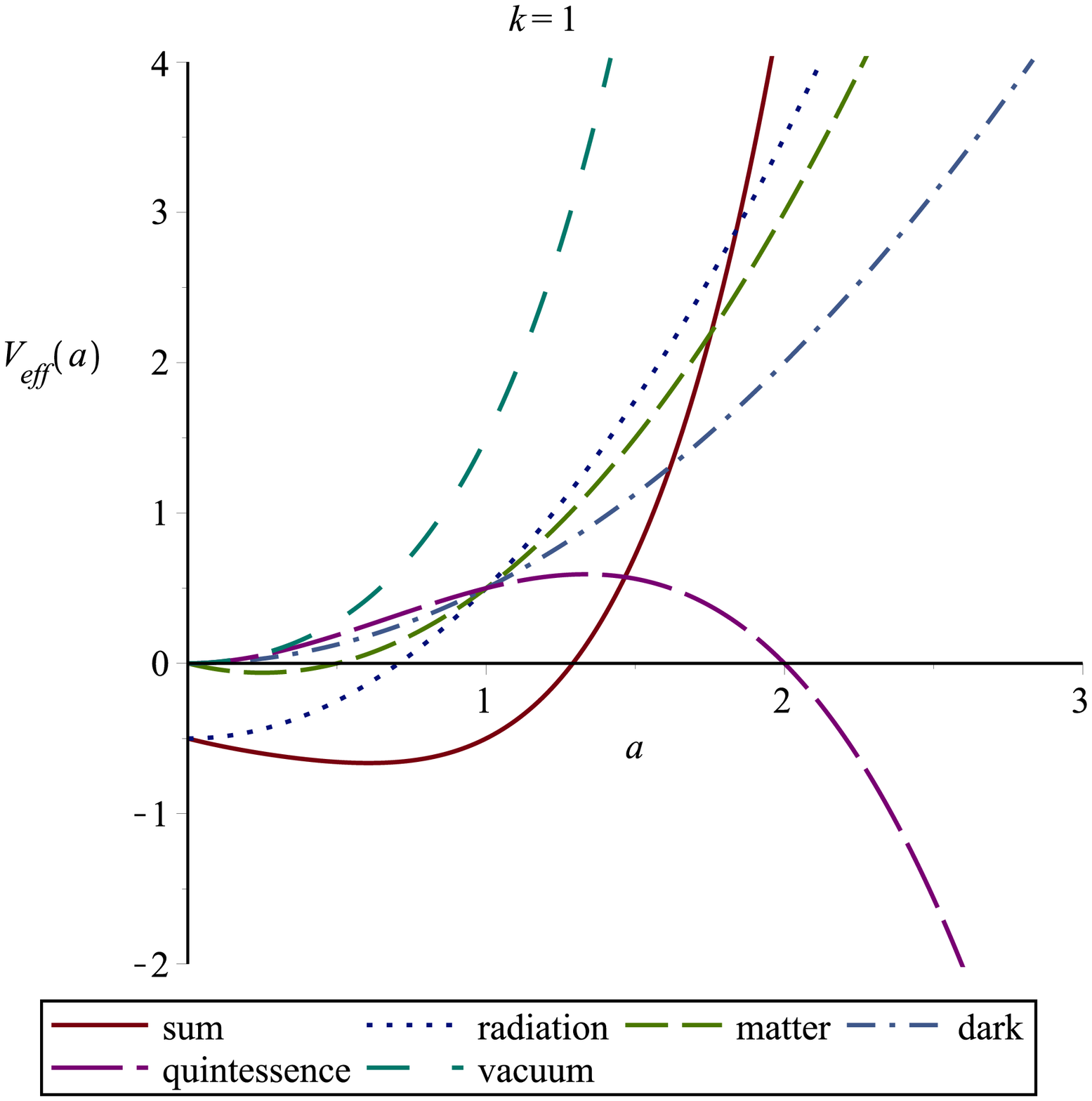}
	\caption{The effective potential energy, $V_{eff}(a)$, for $\Lambda = -|\Lambda|$. We focus on the $k=1$ case and compare with each kind of energy density $\omega$.}
	\label{fig:WDW_Fig6}
\end{figure}

The behaviors of $V_{eff}(a)$ for all cases are shown in Figs.~\ref{fig:WDW_Fig1}-\ref{fig:WDW_Fig6}, for positive and negative values of the cosmological constant.

In what follows we will solve Eq.~(\ref{eq:WDE_mov_1}) and show that this solution is given in terms of the triconfluent Heun function, independently of the kind of energy. Therefore, instead of solve Eq.~(\ref{eq:WDE_mov_1}) separately for each value of the state parameter $\omega$, we will do this in a general way, and thus the solutions will be valid for any value of $\omega$, namely, for all kinds of energy that we are considering. 

Doing an appropriate redefinition of constant, we can write the Wheeler-DeWitt equation in the FRW universe as
\begin{equation}
\frac{d^{2}\Psi}{da^{2}}+\left(B_{0}+B_{1}a+B_{2}a^{2}+B_{3}a^{3}+B_{4}a^{4}\right)\Psi=0,
\label{eq:generalized_WDW_equation_FRW_universe}
\end{equation}
where the coefficients $B_{0}$, $B_{1}$, $B_{2}$, $B_{3}$, and $B_{4}$ are given by
\begin{equation}
B_{0}=\frac{6\pi^{2}c^{2}}{\hbar^{2}G}A_{r},
\label{eq:B0_FRW_universe}
\end{equation}
\begin{equation}
B_{1}=\frac{6\pi^{2}c^{2}}{\hbar^{2}G}A_{m},
\label{eq:B1_FRW_universe}
\end{equation}
\begin{equation}
B_{2}=\frac{6\pi^{2}c^{2}}{\hbar^{2}G}A_{d}-\frac{9\pi c^{6}}{4\hbar^{2}G^{2}}k,
\label{eq:B2_FRW_universe}
\end{equation}
\begin{equation}
B_{3}=\frac{6\pi^{2}c^{2}}{\hbar^{2}G}A_{q},
\label{eq:B3_FRW_universe}
\end{equation}
\begin{equation}
B_{4}=\frac{6\pi^{2}c^{6}}{8\pi\hbar^{2}G^{2}}\Lambda.
\label{eq:B4_FRW_universe}
\end{equation}
Now, let us define a new variable, $x$, such that
\begin{equation}
x=\tau(a+\xi),
\label{eq:x_FRW_universe}
\end{equation}
where the parameters $\tau$ and $\xi$ are given by
\begin{equation}
\tau=\left(-\frac{4B_{4}}{9}\right)^{1/6},\quad \xi=\frac{B_{3}}{4B_{4}}.
\label{eq:tau_FRW_universe}
\end{equation}
Note that the variable $x$ is not necessarily real. In fact, this is not a problem because the special functions that we will use to solve the Wheeler-DeWitt equation can be defined for real parameters as well as for complex ones. Thus, with this new variable, we can write Eq.~(\ref{eq:generalized_WDW_equation_FRW_universe}) as
\begin{equation}
\frac{d^{2}\Psi}{dx^{2}}+\left(b_{0}+b_{1}x+b_{2}x^{2}-\frac{9}{4}x^{4}\right)\Psi=0,
\label{eq:generalized_motion_FRW_universe}
\end{equation}
where the coefficients $b_{0}$, $b_{1}$, and $b_{2}$ are given by
\begin{equation}
b_{0}=\frac{B_{0}-B_{1} \xi +B_{2} \xi ^2-B_{3} \xi ^3+B_{4} \xi ^4}{\tau ^2},
\label{eq:b0_FRW_universe}
\end{equation}
\begin{equation}
b_{1}=\frac{B_{1}-2 B_{2} \xi +3 B_{3} \xi ^2-4 B_{4} \xi ^3}{\tau ^3},
\label{eq:b1_FRW_universe}
\end{equation}
\begin{equation}
b_{2}=\frac{B_{2}-3 B_{3} \xi +6 B_{4} \xi ^2}{\tau ^4}.
\label{eq:b2_FRW_universe}
\end{equation}

Equation (\ref{eq:generalized_motion_FRW_universe}) is similar to the triconfluent Heun equation, which is a particular case of a second order linear differential equation with four singularities, called Heun equation. The canonical form of the triconfluent Heun equation reads as \cite{Ronveaux:1995}
\begin{equation}
\frac{d^{2}y(x)}{dx^{2}}-(\gamma+3x^{2})\frac{dy(x)}{dx}+[\alpha+(\beta-3)x]y(x)=0,
\label{eq:Triconfluent_Heun_Canonical}
\end{equation}
where $y(x)=\mbox{HeunT}(\alpha,\beta,\gamma;x)$ is the triconfluent Heun function. By using the approach described in \cite{AnnPhys.350.14}, we can write Eq.~(\ref{eq:Triconfluent_Heun_Canonical}) in the normal form as
\begin{equation}
\frac{d^{2}Y}{dx^{2}}+\left(\alpha-\frac{1}{4}\gamma^{2}+\beta x-\frac{3}{2}\gamma x^{2}-\frac{9}{4} x^{4}\right)Y=0,
\label{eq:Triconfluent_Heun_normal}
\end{equation}
where $Y(x)=\mbox{e}^{-\frac{1}{2}(x^{3}+\gamma x)}y(x)$. Comparing Eqs.~(\ref{eq:generalized_motion_FRW_universe}) and (\ref{eq:Triconfluent_Heun_normal}), we conclude that the Wheeler-DeWitt equation in the FRW universe, given by Eq.~(\ref{eq:generalized_motion_FRW_universe}) for an arbitrary $\omega$, is algebraically analogous to Eq.~(\ref{eq:Triconfluent_Heun_normal}). Therefore, its exact solution is given in terms of the triconfluent Heun functions and can be written as
\begin{equation}
\Psi(x)=C_{1}\ \mbox{e}^{-\frac{1}{2}(x^{3}+\gamma x)}\ \mbox{HeunT}(\alpha,\beta,\gamma;x),
\label{eq:psi_HeunT_FRW_universe}
\end{equation}
where $C_{1}$ is a constant to be determined, and the parameters $\alpha$, $\beta$, and $\gamma$ are identified as
\begin{equation}
\alpha=b_{0}+\frac{1}{9}b_{2}^{2},\quad \beta=b_{1},\quad \gamma=-\frac{2}{3}b_{2}.
\label{eq:alpha_FRW_universe}
\end{equation}
Thus, the general exact solution of the Wheeler-DeWitt equation in the FRW universe, valid for different kinds of energy, is given in terms of the triconfluent Heun function. The complete set of solutions for each particular value of $\omega$ is summarized in Table \ref{tab:parameters}.

\begin{table}[ph]
\tbl{The parameters $\alpha$, $\beta$, $\gamma$ for the triconfluent Heun function for different values of $\omega$.}
		{\begin{tabular}{ccccc} \toprule
			$\omega$       & $x$            & $\alpha$ & $\beta$ & $\gamma$ \\ \colrule
			0              & $\tau a$       & $\frac{27 c^4 k^2}{16} \sqrt[3]{\frac{3\pi^{2}}{G^{4} \Lambda^{4} \hbar^{4}}}$ & $\frac{6 A_{m}}{c \hbar }\sqrt{-\frac{3\pi^{3}}{\Lambda}}$ & $\frac{3 c^2 k}{2} \sqrt[3]{\frac{9\pi}{G^{2} \Lambda^{2} \hbar^{2}}}$ \\
			$\frac{1}{3}$  & $\tau a$       & $\frac{3 (9 c^4 k^2-32 \pi  A_{r} G \Lambda)}{16}
			\sqrt[3]{\frac{3\pi^{2}}{G^{4} \Lambda^{4} \hbar^{4}}}$ & $0$ & $\frac{3 c^2 k}{2}\sqrt[3]{\frac{9\pi}{G^{2} \Lambda^{2} \hbar^{2}}}$ \\
			-1             & $\tau a$       & $\frac{27 c^{12} k^2}{16}\sqrt[3]{\frac{3 \pi^{2}}{c^{8} G^{4} \hbar^{4}(8 \pi A_{v} G+c^{4} \Lambda)^{4}}}$ & $0$ & $\frac{3 c^{6} k}{2}\sqrt[3]{\frac{9 \pi}{c^{4} G^{2} \hbar^{2}(8 \pi A_{v} G+c^{4} \Lambda)^{2}}}$ \\
			-$\frac{1}{3}$ & $\tau a$       & $\frac{3 (3 c^4 k-8 \pi  A_{d} G)^2}{16 c^4} \sqrt[3]{\frac{3\pi^{2}}{G^{4} \Lambda^{4} \hbar^{4}}}$ & $0$ & $\frac{3 c^4 k-8 \pi  A_{d} G}{2 c^2}
			\sqrt[3]{\frac{9\pi}{G^{2} \Lambda^{2} \hbar^{2}}}$ \\
			-$\frac{2}{3}$ & $\tau (a+\xi)$ & $\frac{9 (256 \pi ^4 A_{q}^4 G^4+64 \pi ^2 A_{q}^2 c^8 G^2 k \Lambda +3 c^{16} k^2 \Lambda ^2)}{16 c^{12} \Lambda ^2 (\frac{9\pi}{G^{2} \Lambda^{2} \hbar^{2}})^{-1/3}}$ & $\frac{3 (16 \pi ^4 A_{q}^3 G^2+3 \pi ^2 A_{q} c^8 k \Lambda )}{c^9 \hbar (-\frac{3}{\pi \Lambda^{5}})^{-1/2}}$ & $-\frac{3 G \hbar  \left(8 \pi ^2 A_{q}^2 G^2+c^8 k \Lambda \right)}{2 c^6 (-\frac{9\pi}{G^{5} \Lambda^{5} \hbar^{5}})^{-1/3}}$ \\ \botrule
		\end{tabular}
	\label{tab:parameters}}
\end{table}

It is worth commenting that in Fig.~(\ref{fig:WDW_Fig1})-(\ref{fig:WDW_Fig6}), we did not taken into account the possible initial values of the energy densities, because we decided to focus only on the explicitly dependence of the effective potential $V_{eff}(a)$ with the scale factor $a$, as shown in Eq.~(\ref{eq:WDW_effective_potential_energy}). To do this, we fixed, in fact, two parameters, namely, the cosmological constant $\Lambda$ and the energy densities $A_{\omega}$, choosing arbitrarily, the values $1$ and $1/2$, for these two quantities, respectively.
%
%
\section{Dynamical interpretation}\label{Sec.III}
In this section we follow the technique developed by He \textit{et al}. \cite{PhysLettB.748.361}, which is based on the method used by Vilenkin \cite{PhysRevD.33.3560} to analyze the dynamical interpretation of the wave function of the universe using the approach which considers the Wheeler-DeWitt equation in the minisuperspace. Thus, let us consider the exact solution of the Wheeler-DeWitt equation in the FRW universe to study the dynamical interpretation of the wave function and discuss the boundary conditions.

As in Eq.~(\ref{eq:generalized_WDW_equation_FRW_universe}) there is only one variable, namely, the scale factor $a$, and thus we can suppose that the solution of the Wheeler-DeWitt equation, $\Psi(a)$, can be written, formally, as
\begin{equation}
\Psi(a)=F(a)\ \mbox{e}^{iS(a)},
\label{eq:psi_expansion}
\end{equation}
where $F$ e $S$ are real functions, in general. Note that these functions will not be used directly in the scenario under consideration, due to the fact that we will consider the conserved probability current, instead, and the Hamilton-Jacobi formulation of the quantum mechanics to write the relations between these functions, the action and the canonical momentum, as explicitly presented along the text. Thus, from Eq.~(\ref{eq:psi_expansion}), we conclude that the square modulus of the wave function of the universe is given by
\begin{equation}
|\Psi(a)|^{2}=F^{2}(a).
\label{eq:square_modulus_FRW_universe}
\end{equation}
Otherwise, the conserved probability current density can be written as \cite{Bransden:2000}
\begin{equation}
j^{a}=\frac{i\hbar}{2\mu}[\Psi^{*}(\partial_{a}\Psi)-\Psi(\partial_{a}\Psi^{*})],
\label{eq:current_density_FRW_universe}
\end{equation}
where $\mu$ is the mass. This expression was constructed in order to guarantee the validity of the continuity equation, namely,
\begin{equation}
\partial_{a}j^{a}=0.
\label{eq:conserved_FRW_universe}
\end{equation}
Substituting Eq.~(\ref{eq:psi_expansion}) into Eq.~(\ref{eq:current_density_FRW_universe}), we obtain
\begin{equation}
j^{a}=-\frac{\hbar}{\mu}F^{2}\frac{\partial S}{\partial a}.
\label{eq:j_1_FRW_universe}
\end{equation}
On the other hand, integrating Eq.~(\ref{eq:conserved_FRW_universe}) we get
\begin{equation}
j^{a}=C_{0},
\label{eq:j_2_FRW_universe}
\end{equation}
where $C_{0}$ is a constant. Thus, from Eqs.~(\ref{eq:j_1_FRW_universe}) and (\ref{eq:j_2_FRW_universe}), we have
\begin{equation}
-\frac{\hbar}{\mu}F^{2}\frac{\partial S}{\partial a}=C_{0}.
\label{eq:result_1_FRW_universe}
\end{equation}
Then, we may use the Hamilton-Jacobi formalism of quantum mechanics to write the following relation between the action and the canonical momentum
\begin{equation}
p_{a}=\frac{\partial S}{\partial a}=\frac{\partial L}{\partial \dot{a}}=-\frac{3 \pi c^{2}}{2G}\dot{a}a,
\label{eq:action_momentum_FRW_universe}
\end{equation}
where $L$ is the Lagrangian for the FRW universe, given by \cite{IntJModPhysD.11.527}
\begin{equation}
L=-\frac{3 \pi c^{2}}{4G}a^{3}\left[\left(\frac{\dot{a}}{a}\right)^{2}-\frac{k c^{2}}{a^{2}}+\frac{8 \pi G}{3 c^{2}}(\rho+\rho_{vac})\right].
\label{eq:Lagrangian_WDW}
\end{equation}
Thus, from Eqs.~(\ref{eq:result_1_FRW_universe}) and (\ref{eq:action_momentum_FRW_universe}), we get
\begin{equation}
F^{2}=C_{0}\frac{2 \mu G}{3 \hbar \pi c^{2} \dot{a} a}.
\label{eq:result_2_FRW_universe}
\end{equation}

Now, consider the solution of the Wheeler-DeWitt equation in the $a \gg 1$ limit, which implies that $x \gg 1$. In this case, the triconfluent Heun function has the following asymptotic behavior
\begin{equation}
\mbox{HeunT}(\alpha,\beta,\gamma;x) \sim x^{\frac{\beta}{3}-1}\sum_{\nu \geq 0}a_{\nu}(\alpha,\beta,\gamma)x^{-\nu},
\label{eq:HeunB_infty_FRW_universe}
\end{equation}
where $|\arg x| \leq \frac{\pi}{2}$, and $a_{0}(\alpha,\beta,\gamma)=1$. As a consequence the wave function can be written as
\begin{equation}
\Psi(a) \sim \frac{C_{1}}{a},
\label{eq:psi_R_HeunT_FRW_universe}
\end{equation}
which implies that
\begin{equation}
|\Psi^{2}(a)| = \frac{C_{1}^{2}}{a^{2}} = F^{2},
\label{eq:psi_R_2_HeunT_FRW_universe}
\end{equation}
where we have used the fact that $\beta$ is an imaginary number. Thus, taking into account Eq.~(\ref{eq:result_2_FRW_universe}), we get
\begin{equation}
\frac{da}{a}=\frac{2 \mu G C_{0}}{3 \hbar \pi c^{2} C_{1}^{2}}\ dt.
\label{eq:separated_FRW_universe}
\end{equation}
Integrating Eq.~(\ref{eq:separated_FRW_universe}), we obtain the following asymptotic behavior for the scale factor
\begin{equation}
a(t) \propto \mbox{e}^{t+t_{0}}.
\label{eq:scale_factor_asymptotic}
\end{equation}
From Eq.~(\ref{eq:scale_factor_asymptotic}), we can conclude that the evolution law of the universe obtained in the framework of quantum cosmology, taking into account the classical limit ($a \gg 1$) is completely consistent with the solution of the Friedmann equation when the vacuum energy dominates, which means that the universe will behave according to the energy contained in the vacuum, whatever the form of energy. This result is analogous to the ones that we have obtained in the quantum Newtonian cosmology scenario \cite{JMathPhys.60.102301}, in which case the universe is described using Newtonian mechanics, assuming the cosmological principle and adopting the fact that the universe experiences an expansion. Notice that the present results are independent of the kind of energy and hence it is more general than the ones found in the literature \cite{dInverno:1998}.

Next, consider the solution of the Wheeler-DeWitt equation in the $a \ll 1$ limit, which implies that $x \ll 1$. In this case, the expansion in a power series for all $x$ of the triconfluent Heun function is given by \cite{PhysRevD.94.023511}
\begin{equation}
\mbox{HeunT}(\alpha,\beta,\gamma;x)=\sum_{s \geq 0}u_{s}(\alpha,\beta,\gamma)x^{s},
\label{eq:WDE_Triconfluent_Heun_expansion}
\end{equation}
where $u_{0}=1$. Thus, the wave function given by Eq.~(\ref{eq:psi_HeunT_FRW_universe}) for small scale factor values, namely, $a \ll 1$, can be written as
\begin{equation}
\Psi(a) \sim C_{1},
\label{eq:psi_R_small_HeunT_FRW_universe}
\end{equation}
and thus the squared modulus of the wave function is given by
\begin{equation}
\Psi^{2}(a) = C_{1}^{2}.
\label{eq:psi_R_small_2_HeunT_FRW_universe}
\end{equation}
Now, taking into account once more Eq.~(\ref{eq:result_2_FRW_universe}), we get
\begin{equation}
a\ da=\frac{2 \mu G C_{0}}{3 \hbar \pi c^{2} C_{1}^{2}}\ dt,
\label{eq:separated_small_FRW_universe}
\end{equation}
which tell us that, in this limit,
\begin{equation}
a(t) \sim (t+t_{0})^{\frac{1}{2}}.
\label{eq:evolution_small_FRW_universe}
\end{equation}
The behavior of the scale factor given by Eq.~(\ref{eq:evolution_small_FRW_universe}) means that when the universe was very small, its behavior was determined by radiation. Indeed, at earliest epochs, the matter is relativistic.

At this point, let us call attention to the fact that in order to obtain the results for $a \gg 1$ and $a \ll 1$, we considered the obtained general solution, given by Eq.~(\ref{eq:psi_HeunT_FRW_universe}), which is valid for all kind of energy densities under consideration, taken together simultaneously. The $a \gg 1$ limit was examined by imposing this condition to Eq.~(\ref{eq:psi_HeunT_FRW_universe}). This limit corresponds to the end of the expansion of the universe. Otherwise, the $a \ll 1$ limit corresponds to early stages of the universe. In principle, the scenario in which only some kind of energy densities are considered, as for example, radiation and matter, as a particular case can be investigated. To do this it seems to be better to start from the very beginning with these two kinds of energy densities, due to the mathematical difficulties to manage the solution given in terms of the Heun function, for the general case, and assume that some parameters are equal to zero, in which case a confluent process may appears.
%
%
\section{Tunneling universe}\label{Sec.IV}
\begin{figure}[htbp]
	\centering
		\includegraphics[scale=0.50]{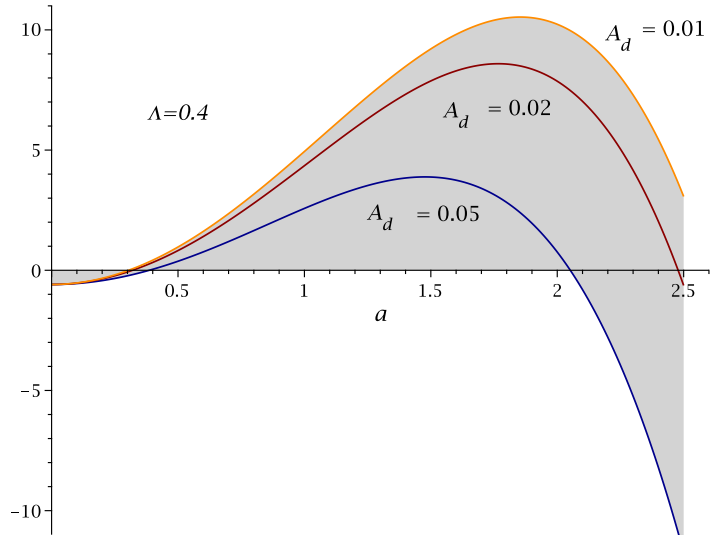}
	\caption{The effective potential energy, $V_{eff}(a)$, for different values of $A_d$. We focus on the positive curvature case, $k=1$.}
	\label{fig:WDW_Fig7}
\end{figure}

In this section, in order to describe the early universe with zero energy in its tunneling phase, we calculate the transmission coefficient through the effective potential barrier, which is based on the method developed by Vilenkin \cite{PhysRevD.37.888} and permits the realization of the phenomenon. For this purpose, the effective potential is given by
\begin{equation}
V_{eff}=-(B_0+B_2a^2+B_4a^4),
\label{A}
\end{equation}
whose behavior is depicted in Fig. \ref{fig:WDW_Fig7}, where $B_2<0$. Notice the absence of the matter as well quintessence in the potential contents of energy, given by Eq.~(\ref{A}). This restriction was assumed in order to simplify our problem. The referred transmission coefficient reads
\begin{equation}
T=\frac{\vec{j}^{out}\cdot\hat{n_o}}{\vec{j}^{in}\cdot\hat{n_i}},
\label{transmission}
\end{equation}
where ${j}^{in,out}_a=i\hbar\{\Psi^{*}(a_{\mp})[\partial_{a}\Psi(a_{\mp})]-\Psi(a_{\mp})[\partial_{a}\Psi^{*}(a_{\mp})]\}$ are the probability current densities calculated at the points where $V_{eff}=0$, relative to the {\it cis}-barrier and {\it trans}-barrier regions, respectively, and $\hat{n}_{i,o}$ are the unit vectors normal to the potential barrier at those points. Such points, which are the roots of $V_{eff}$ function for $a\geq0$, will be designated as $a_-$ and $a_+$, and are given by
\begin{equation}
a_{\pm}= \sqrt{\frac{-B_2\pm\sqrt{B_2^2-4B_0B_4}}{2B_4}},
\label{rootspotential}
\end{equation}
with $B_2<0$. It is worth calling attention to the fact that, without any content of radiation, we cannot have cosmic tunneling. In fact, the content of radiant energy warrants the unit vector $n_i$ to have a non-null component parallel to the $a$-axis in the effective potential graph, which allows the tunneling. The cosmological constant is also necessary, provided that $2\sqrt{B_0B_4}<|B_2|$. On the other hand, the dark energy content must be such that $A_d<\frac{3c^4}{8\pi G}$, since $B_2<0$. Thus, the energy contents of the early universe to be considered are the radiation, dark energy ($\omega=-1/3$) and vacuum energy in the form of a positive cosmological constant $\Lambda$.

Therefore, we can write Eq.(\ref{transmission}) as
\begin{eqnarray}
T & = & \frac{\Psi^{*}(a_+)[\partial_{a}\Psi(a_+)]-\Psi(a_+)[\partial_{a}\Psi^{*}(a_+)]}{\Psi^{*}(a_-)[\partial_{a}\Psi(a_-)]-\Psi(a_-)[\partial_{a}\Psi^{*}(a_-)]}\nonumber\\
&& \times\frac{\sin{[\tan^{-1}{V'_{eff}(a_+)]}}}{\sin{[\tan^{-1}{V'_{eff}(a_-)}]}}.
\label{transmission2}
\end{eqnarray}
Henceforth, we will consider that the universe in the tunneling phase was very small. Thus, inserting Eqs.~(\ref{eq:WDW_effective_potential_energy}), (\ref{eq:psi_HeunT_FRW_universe}) and (\ref{rootspotential}) into Eq.~(\ref{transmission2}), and using the expansion of the triconfluent Heun function up to second order in $a$, $\mbox{HeunT}(\alpha,\gamma,\delta;x) \approx 1-\alpha x^2/2$, as well as the expansion of the $\sin(\arctan{z})$ up to this same order, we arrive at
\begin{equation}
T \approx \left(\frac{a_+}{a_-}\right)e^{-\frac{1}{2}\sqrt{\frac{3B_2^2}{B_4}}(a_+-a_-)}.
\end{equation}
Notice that, for small values of the radiation content, the argument of the decreasing exponential is proportional to both the square root of the effective potential barrier height, $V_{max}\approx B_2^2/4B_4$, and width, $a_+-a_-$, as expected. We point out that as higher as the cosmological constant, much greater is the transmission coefficient, {\it i.e.}, the vacuum energy favors the early cosmic tunneling. Figure \ref{fig:WDW_Fig7} illustrates how larger values of the dark energy, $A_d$, contributes for a greater probability of tunneling, turning smaller and thinner the effective potential barrier. On the other hand, the cosmological constant also contributes to this phenomenon, since a greater value of this quantity decreases $B_4$ and, therefore, increases the transmission coefficient.

%
%
\section{Conclusions}\label{Sec.V}
We generalized some results in the literature, taking into account the description provided by the Wheeler-DeWitt equation, in the sense that we have obtained now the exact solutions for each form of the energy density. The relativistic wave functions are given in terms of the triconfluent Heun functions and hence obey the appropriate boundary conditions.

As we can see from Figs.~\ref{fig:WDW_Fig1}-\ref{fig:WDW_Fig6}, at earliest epochs the behavior of the effective potential is dominated by the energy density of the radiation. Next, there exist a matter domination epoch and then the energy density contained in the vacuum dominates the behavior of the universe. Therefore, our approach is completely consistent with the solutions of the Friedmann equation.

We analyzed the dynamical interpretation for the wave function of the FRW universe. This interpretation gives the behavior of the scale factor at the end of the expansion, which is in accordance with the correspondence principle applied to quantum cosmology, namely, we recover the classical results. Furthermore, when the universe was very small, the quantum effects on the scale factor are dominated by the energy density of the relativistic matter.

We also calculated the exact expression for the probability of the early universe to tunnel through the effective potential barrier, considering a positive curvature, $k=1$. The computed transmission coefficient acquires such a meaning, associated to the quantum tunneling of the universe, in the context of the semiclassical discussion made in the previous section. For this, we have considered low values of radiation ($\omega=1/3$), as well as the presence of contents of dark energy ($\omega=-1/3$) and vacuum energy. In this last case, we considered a positive cosmological constant. These energy contents yield an effective potential barrier with the shape required to the occurrence of the tunneling. By taking into account a very small size of the universe in its early stages, we found that the tunneling probability depends on a decreasing exponential whose argument is proportional to both the square root of the height and to the width of the effective potential barrier, for low contents of radiant energy. The obtained results pointed out that higher values of the vacuum and dark energy densities favor the quantum tunneling through the barrier. It is worth to mention that some cosmological models require large values for the cosmological constant at the early universe \cite{Lucca}.

Note that we have considered in this work an ordering factor $p=0$, so that solutions to the Wheeler-DeWitt equation, valid for all kinds of energy, are analytically tractable. However, it would be interesting to extend our analysis to a general case, where the ordering factor is taking into account.

Finally, it is worth emphasizing that, in principle, we can get information about the role played by the initial values of the energy densities on the evolution of the universe, by considering the obtained solution given by Eq.~(\ref{eq:psi_HeunT_FRW_universe}) and the parameters in the argument of the wave function defined in Table \ref{tab:parameters}, in terms of the ones associated with the energy densities. Doing this, we would be able to know, quantitatively, as well qualitatively, how the initial values of the energy densities influence the history of the universe. In practice, however, we expect that there is some difficulties from the mathematical point of view related to the complexity of the obtained solution. In fact, unfortunately, there are still some gaps in the theory of the Heun's differential equations.
%
%
\section*{Acknowledgments}
H.S.V. was funded through the CNPq research Project No. 150640/2018-8 and also by the Coordena\c c\~{a}o de A\-per\-fei\-\c co\-a\-men\-to de Pessoal de N\'{i}vel Superior - Brasil (CAPES) - Finance Code 001. V.B.B. is partially supported from CNPq Project No. 305835/2016-5. M.S.C. is partially supported from CNPq Project Numbers 312251/2015-7, 433168/2016-1 and 314183/2018-3. The authors would like to thank FUNCAP for partial financial support under the grant PRONEM PNE-0112-00085.01.00/16. The authors also would like to thank Professor Daniel Vanzella for the fruitful discussions.
%
%

%
%
\end{document}